\documentclass[sigconf]{acmart}

\AtBeginDocument{%
  }

\copyrightyear{2025}
\acmYear{2025}
\setcopyright{rightsretained}
\acmConference[CHI EA '25]{Extended Abstracts of the CHI Conference on Human Factors in Computing Systems}{April 26-May 1, 2025}{Yokohama, Japan}
\acmBooktitle{Extended Abstracts of the CHI Conference on Human Factors in Computing Systems (CHI EA '25), April 26-May 1, 2025, Yokohama, Japan}
\acmDOI{10.1145/3706599.3716214}
\acmISBN{979-8-4007-1395-8/25/04}

\begin{document}

\title{Unlimited Editions}
\subtitle{Documenting Human Style in AI Art Generation}

\author{Alex Leitch}
\email{aleitch@umd.edu}
\affiliation{%
  \institution{College of Information}
  \institution{University of Maryland}
  \city{College Park}
  \state{Maryland}
  \country{USA}}

\author{Celia Chen}
\email{clichen@umd.edu}
\affiliation{%
  \institution{College of Information}
  \institution{University of Maryland}
  \city{College Park}
  \state{Maryland}
  \country{USA}}

\begin{abstract}
As AI art generation becomes increasingly sophisticated, HCI research has focused primarily on questions of detection, authenticity, and automation. This paper argues that such approaches fundamentally misunderstand how artistic value emerges from the concerns that drive human image production. Through examination of historical precedents, we demonstrate that artistic style is not only visual appearance but the resolution of creative struggle, as artists wrestle with influence and technical constraints to develop unique ways of seeing. Current AI systems flatten these human choices into reproducible patterns without preserving their provenance. We propose that HCI's role lies not only in perfecting visual output, but in developing means to document the origins and evolution of artistic style as it appears within generated visual traces. This reframing suggests new technical directions for HCI research in generative AI, focused on automatic documentation of stylistic lineage and creative choice rather than simple reproduction of aesthetic effects.
\end{abstract}



\begin{teaserfigure}
  \includegraphics[width=\textwidth]{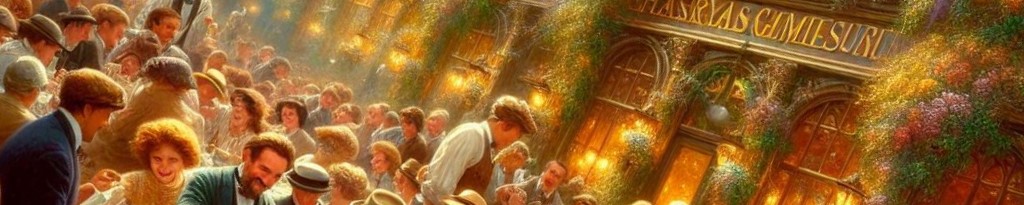}
  \caption{"Thomas Kinkade painter of light image Renoir people at lunch", Image Creator in Bing}
  \Description{AI-generated image combining stylistic elements of Thomas Kinkade and Pierre-Auguste Renoir, created using Bing Image Creator.}
  \label{fig:teaser}
\end{teaserfigure}

\received{5 December 2024}
\received[accepted]{30 January 2025}

\maketitle

\section{Introduction}
In 1893, at the Chicago World’s Fair, visitors marveled at mass-produced copies of masterworks - mechanical reproductions that promised to democratize art by making it accessible to all \cite{elam2022signs}. A century later, we face a similar moment of technological disruption with AI image generation. Just as those early reproductions sparked fierce debates about authenticity and meaning in art, today's AI systems like Midjourney and DALL-E raise profound questions about creativity, culture, and what we value in artistic expression. Recent experiments suggest many viewers cannot distinguish AI-generated art from human-made works \cite{alexander2024ai}, but this reveals less about AI capabilities than it does about our own cultural blind spots. When American audiences regularly confuse Japanese and Chinese classical art \cite{eleta2012study}, or when casual viewers fail to spot the telltale marks of mass reproduction, we're not witnessing the triumph of technology - we're seeing the limits of untrained perception. This superficial approach to evaluating art poses significant risks. As AI systems become increasingly capable of mimicking visual styles, we risk reducing centuries of artistic innovation and cultural meaning to mere surface-level features that can be computationally reproduced.

Today's AI systems differ from historic reproduction systems, such as lithography. They do not simply copy existing works, but rather learn to generate endless variations in acquired styles - effectively creating what Dr. Kate Compton calls a "Bach faucet" \cite{compton2023declaring} of infinite content in a previously-valued cultural mode. This very infinity then renders the output worthless. While much recent attention has focused on whether machines can create "real" art or if viewers can distinguish AI-generated from human-created works, these discussions miss how art actually functions as a store of value in human society. The art world has long maintained complex systems for authenticating and valuing creative work precisely because human artistic production is finite. While researchers meticulously measure whether humans can detect AI-generated art or document the intricacies of prompt engineering, we are missing a more provocative truth: every human artist will die. Their portfolios are ultimately limited. Far from a weakness, this limitation forms the value basis of creative work as art.

Rather than attempting to solve questions of AI creativity or authenticity, we propose that HCI researchers should focus on developing systems for documenting and attributing the human origins of AI style transfer - not just as a matter of credit or copyright, but as a fundamental acknowledgment that each human artist's creative style represents something unrepeatable: the output of a singular finite life. This reframing suggests new technical approaches focused not on generation but on preservation and attribution of human creative development.

Through examination of both contemporary cases and the historical example of the Impressionist movement, we demonstrate why the bounded nature of human creative production - shaped by individual experience, artistic lineage, and conscious creative choices - transforms style from aesthetic surface into valuable cultural artifact. We conclude by suggesting that HCI's role in the future of AI art lies not in perfecting generation or detecting fakery, but in developing systems that document and preserve the human struggle of artistic work - the specific choices, influences, and creative resolutions that shape each artist's unique way of seeing. As style becomes infinitely reproducible through AI systems, maintaining these connections to human creative development becomes not just important but essential.

\section{Background and Related Work}
\subsection{Historical Context}
The late 19th century witnessed a seismic shift in the art world, as the emergence of accessible photographic reproduction kicked off a related rise in avante-garde painting movements responding to the new technology and new social circumstances of the Industrial Revolution. From this period came the Impressionist movement, one of the most-reproduced stylistic modes in contemporary imagemaking, both generative and human. The Impressionists revolutionized painting techniques and challenged then-accepted understandings of the role of artistic representation, even as mechanical reproductions of fine art challenged the role of audience appreciation \cite{spencer2016impressionism}.

\subsubsection{Technological innovations and their impact}
The invention of the collapsible metal paint tube by American portrait painter John G. Rand in 1841 revolutionized the practice of painting. The tube portioned pre-mixed paint and preserved it against drying in the air. This allowed artists to easily transport bright, consistent colors, enabling them to paint outdoors (en plein air) and capture the mutable effect of light and atmosphere directly from nature \cite{greenspan2013never}. The portability and convenience of paint tubes were crucial to the Impressionists' ability to work quickly and spontaneously, capturing brief moments of time in the permanence of fresh new oil.

 Although photography is often described as directly representative, successful image-making involves a collection of novel technical skills and choices \cite{sontag1977photography} in what the camera should see \cite{sauter2022seeing}. The arrival of albumen print photography in the mid-19th century had a complex relationship with painting. While popular culture holds an idea of photography in the 1800s as static, taking hours, the wet-colloidal process introduced in 1851 was by the mid-1860s overwhelmingly popular for common photographs \cite{scientific1915invention}. It is difficult to express the scale of change this presented to accessible representations of everyday life. The sudden availability of relatively reliable, portable image reproduction led to a dynamic relationship between painters and photography. As photographers mastered precise depiction of figurative images, painters found themselves free to pursue two previously challenging extremes: perfectly casual, true-to-life scenes, and radically abstract forms that entirely departed from literal representation. The cropped compositions and unusual angles often seen in Impressionist paintings reflect the influence of photographic techniques \cite{fletcher2012rise}. 

\subsubsection{The Impressionist revolution}
Starting in Paris circa 1860, the Impressionist movement marked a significant departure from established artistic norms of the period. Led by pioneers such as Claude Monet, Pierre-Auguste Renoir, and Camille Pissarro, this group of painters changed how representational painting worked within the French academy. Their technical concern was the newly-possible capture of ephemeral effects of light and atmosphere in everyday, often outdoor settings. This was a fundamental shift in subject matter, allowing audiences to engage with work that showed them their own lives, rather than idealized historic scenes created in studio settings from hand-mixed linseed oil and pigment. This radical approach to painting not only transformed the art world of the late 19th century, but also laid the groundwork for many of the avant-garde movements that would follow in the 20th century \cite{minneapolis_nd}.

Impressionist paintings are distinguished by innovative painting techniques that distinguish their work from conventional art practices of the time. They employ loose, visible brushstrokes and apply pure, unmixed colors side by side on the canvas, allowing the viewer's eye to blend the hues optically \cite{callen2000art}. This causes interesting effects in the visual cortex, resulting in a sense of vibrant immediacy in their paintings. In order to capture the shift in light and atmosphere throughout the day, they were some of the earliest painters to regularly work outdoors, from life \cite{spencer2016impressionism}, leading to a sense of spontaneity and immediacy in their work.

The Impressionists were controversial at the time, but their legacy and impact is such that they are now some of the most-reproduced images of Western fine art in the world. Their works reliably fetch millions at auction \cite{kinsella2024most}, and their touring exhibitions - particularly Monet and Renoir - are key sales drivers for major museum and gallery tickets \cite{lauter2024impressionism}. The subject matter and style of the Impressionists have also become routine to the visual vernacular of social media sharing sites such as Instagram.

\subsection{Commercial Art Reproduction and Its Consequences}
The 19th century, when the Impressionists were working, witnessed a significant expansion in the commercial art reproduction industry, driven by technological advancements and increasing consumer demand \cite{verhoogt2019art}. The rise of mass reproduction techniques such as lithography and photography abruptly made fine art available to a broad audience.

Art dealer-publishers capitalized on this trend by employing innovative marketing strategies, often leveraging nationalist ideology to promote their reproductions and appeal to patriotic sentiments. This commercialization of art reached new heights with the use of fine art for advertising purposes, exemplified by the appropriation of Millais' "Bubbles" painting for Pears soap advertisements, and a massive expansion in illustrative lithographic posters generally \cite{marzio1971lithography}. This signaled a significant shift in the relationship between art and commerce, establishing patterns of commercial reproduction and value that would shape the art market for the next century.

The parallel between commercial reproduction houses of the 19th century and modern AI art systems is telling. Both technologies promised to democratize art by making it more accessible, but in doing so, both risked reducing unique artistic voices to reproducible formulas. However, where historical reproduction maintained a clear distinction between original and copy, AI systems blur this line by generating new works that appear authentic while lacking the finite human context that gives art its cultural and market value.

\subsection{Contemporary AI Art Generation}
Prior work examining AI art generation has primarily focused on perception and evaluation studies comparing fully AI-generated versus human-generated art, with an emphasis on whether viewers can distinguish between the two sources. This focus on detection and authenticity echoes Walter Benjamin's prophetic 1935 observations about mechanical reproduction - but where Benjamin saw reproduction threatening the "aura" of original artworks \cite{benjamin1935work}, AI systems present an even more fundamental challenge by generating endless variations that appear authentic. Studies indicate that people have difficulty in correctly identifying AI-generated art, with detection accuracy only marginally above chance. For example, Ragot \cite{ragot2020ai} found humans could detect machine-generated artworks with just 56\% accuracy. Beyond mere detection, research has examined how disclosure of AI involvement affects viewer perception and appreciation. \cite{chamberlain2018putting} and \cite{kobis2021artificial} found that people tend to rate AI-generated art lower once its origins are revealed, displaying an 'algorithm aversion' particularly strong in subjective, creative domains.

Beyond binary comparisons of human versus AI art, research has shown that cultural context and interpretation significantly influence how artwork is perceived and categorized. Eleta and Golbeck \cite{eleta2012study} found that when tagging artworks, American viewers frequently confused Japanese and Chinese art, highlighting how cultural knowledge shapes art interpretation. This cultural mediation extends to AI art reception - Latikka \cite{latikka2023ai} demonstrated that attitudes toward AI in art vary significantly based on broader cultural frameworks and relationships with technology. More fundamentally, Bellaiche \cite{bellaiche2023humans} found that viewers ascribe significantly less agency and experiential capacity to AI systems compared to human artists, with this reduced mind perception correlating with lower appreciation of AI-generated art. This bias appears especially pronounced when evaluating artwork's deeper meaning and worth, rather than just surface aesthetic qualities. These findings suggest that reactions to AI involvement in art creation are shaped not only by viewers' cultural frameworks but also by deeper assumptions about creativity, agency, and artistic authenticity.

Recent work has begun examining human-AI collaboration rather than purely automated art generation. Researchers have found that viewers generally respond more positively to human-machine collaborations compared to fully automated AI art \cite{hitsuwari2023does}\cite{kern2022humans}. This preference may relate to Xu's \cite{xu2020machine} finding that revealing AI involvement negatively affects not only aesthetic appreciation, but also perceptions of authenticity, creativity, and emotional resonance compared to human-created artworks. However, these studies have typically treated collaboration as a single condition rather than examining how different types or stages of human-AI interaction might affect perception. And while computational approaches to analyzing art have grown more sophisticated \cite{cetinic2022understanding}, there remains limited research on how viewers perceive and value different forms of algorithmic assistance in the creative process.

The practice of prompt engineering - writing textual inputs for generative AI models - has emerged as another key area of study \cite{liu2022design}. Studies have shown that prompt engineering requires both technical knowledge and creative expertise, with practitioners often engaging in systematic experimentation through trial and error to achieve desired results \cite{oppenlaender2022creativity}. While commercial tools aim to make AI art generation accessible to novices, research indicates a gap between casual users and expert practitioners who understand how to effectively use prompt modifiers and other advanced techniques \cite{oppenlaender2023perceptions}. This raises questions about whether prompt engineering represents an intuitive skill that people can apply immediately, or whether it requires specific training and expertise to master.

The interaction between human skill and AI capabilities connects directly to broader questions about creative agency and artistic authenticity. As Hong and Curran \cite{hong2019does} found in their survey-experiment, human-created artworks and AI-created artworks were not judged to be equivalent in their artistic value, and individuals' preexisting schemas about AI's ability to create art significantly influenced their evaluations. This suggests that the technical and creative elements of prompt engineering cannot be easily separated from persistent questions about human versus automated creativity.

\subsection{Evolving Artist-AI Dynamics}
More recent work has begun examining how artists are actively negotiating their relationship with AI tools, moving beyond simple human-versus-machine comparisons. As Don-Yehiya \cite{don2023human} found in studying Midjourney users, many artists are developing sophisticated practices that blend human creative intent with AI capabilities. However, tensions persist around questions of artistic control, creative ownership, and authentic expression. This is clearly evident in debates around training data practices and the appropriation of artists' styles without consent \cite{elam2022signs}.

While some artists view AI as simply another digital tool in their creative arsenal, others are exploring more radical possibilities for human-AI co-creation. For instance, Kim \cite{kim2024enhancing} found that artists develop different strategies for "finding the best query" when working with AI image generation systems, with expertise developing through iterative experimentation rather than intuitive understanding. This suggests that meaningful human-AI art creation may require new forms of literacy and practice that go beyond traditional artistic training.

These emerging artist-AI practices raise fundamental questions about creative process and authorship. These questions have been present for some time now, hovering around early-stage work with GAN systems, such as Jake Elwes’ “Machine Learning Porn” \cite{elwes2016machine}, “the Zizi Show” \cite{goldstein2023ai}, or “the Portrait of Edouard de Bellamy” by the art collective Obvious - which sold for more than \$432,000 USD at auction \cite{schneider2018first}.

The evolution of these dynamics is visible in how artists approach prompt engineering. Oppenlaender \cite{oppenlaender2022creativity} documents how practitioners in AI art communities have developed sophisticated techniques for controlling generative outputs, including the use of specific keywords and modifiers that act as a specialized artistic vocabulary. However, this emerging practice raises questions about accessibility and democratization - whether AI art tools are truly making creativity easier, or creating new barriers to entry through technical complexity - and subsequently, a market that may not require a painter at all.

\subsection{Structural Power and Platform Politics}
The development of AI art tools cannot be separated from broader questions of power and control in technological systems. As Elam \cite{elam2022signs} argues, many AI art platforms rely on problematic assumptions about race, culture, and creativity that reflect and reinforce existing power structures \cite{aaron2024ai}. These issues manifest through multiple intersecting mechanisms in current AI art systems. Training data practices frequently appropriate artists' work without their consent or compensation, while platform designs tend to privilege certain aesthetic traditions and cultural perspectives over others. The economic models underlying these systems potentially undermine artists' ability to sustain their creative practice, while their technical infrastructures often encode problematic assumptions about identity and creativity.

Recent work by artist-technologists, particularly those from marginalized communities, has begun challenging these structural issues. Rather than accepting commercial imperatives of "personalization" and "frictionlessness," these practitioners are exploring alternative approaches that recognize identity categories like race and gender not as fixed data points but as dynamic social processes \cite{elam2022signs}. This tension between commercial AI art platforms and alternative artistic practices points to larger questions about the future of creative production. As Zylinska \cite{zylinska2020ai} notes, the challenge is not simply about preventing harm to artists, but about imagining and creating different possibilities for human-AI creative collaboration that don't reproduce existing power inequities.

\subsection{Methodological and Ethical Considerations}
Traditional HCI methodologies focusing on usability metrics or preference ratings cannot adequately capture the deeper questions of artistic authenticity and cultural value. As Sterling \cite{sterling2019work} argues, we need frameworks that can address not just technical performance but the political and cultural tensions inherent in using data platforms to reflect on our social world.

These methodological limitations reflect deeper assumptions about creativity and value. Current practices in AI art generation risk perpetuating what Elam \cite{elam2022signs} terms "epistemic forgeries" - assumptions about knowledge and creativity that help AI function as a technology of power. These include presumptions about universal standards of aesthetics and reductive approaches to cultural expression. Moving beyond these limitations requires us to reconsider fundamental questions about how art comes to be recognized and valued in human society.

\section{Critical Analysis}
Current approaches to studying AI art systems often frame issues narrowly in terms of technical performance or user acceptance, missing more interesting questions about how art comes to be art at all. Recent work highlights that, barring obvious errors, human coders can often not classify something accurately as generated or not, as human or not, and that they may even en-masse gently prefer a generated work. To discover that a system intended to fool the untrained eye has succeeded in doing so is not terribly surprising, but it is revealing; there is an obvious gap in generative image research as to what counts as “art” at all. The nature of human production is that it is limited, both chronologically and in volume. For any great name - any artist who might be named in a generative capture set, such as the list that served as the basis of the artist lawsuit over the training of Midjourney and LAION-5B \cite{goldstein2023b} - the number of produced works is a strictly finite thing. Though it may be mathematically possible to create a new “Bach” sonata \cite{ball2011music}, the Goldberg Variations cannot be changed. They are a set, limited in time to the lifespan of the performer and his specific abilities as they relate to a yet-more-limited set of musical notation by an acknowledged master. 

This produces an interesting problem in the discipline of art history, which is by no means a space lacking in value or free capital. Especially for dead artists - famously the most costly \cite{morris2019dos}\cite{ekelund2000death} - there can only be so many objects recognized as a Renoir. The fine art market has long been a hub to move money around the world, so long as it is attached to appropriate objects - and deciding which objects are appropriate is a large part of the business \cite{prinz2015success}. It does not matter if someone can exactly reproduce the style of the painting; it does not signify that a new painting, identical to the technical and chronological constraints of production, emerges \cite{waldemar2015rodin}. Because these objects represent the motion of captured capital, they are tightly restricted within a system of recognition and reporting. In the West, this system is based in Paris, New York, and the major museums that produce what is recognized as Western culture. Although there is no central registry, major artists are well-documented in their catalog raisonné - a comprehensive, annotated listing of all known artworks by an artist - and the provenance - the specific location and ownership history - of any given piece of work helps to set its value \cite{barham2015worth}.

This scarcity is the current mechanism dictating whether or not an object is art, in the sense that it can be used to capture and transfer value. Art-as-artefact is not only the product of individual creative choices. Rather, it is the end result of a process of international trade, mediated by auction houses such as Sotheby’s and Christie’s across the globe. In this system, the skill of exact replication is assumed, and is often assigned to low-paid technicians \cite{cronin2012collaboration} - sometimes these technicians are even computational systems. In order to attain the status of art, the object-as-artefact must first emerge from a system of gallerists, collectors, and dealers. The system begins at market fairs, such as Art Basel, Frieze, and for digital works Transmediale, where value is negotiated for early-career artists by dealers and acquisition curators \cite{morgner2014evolution}.  The results of art auctions are important enough to merit their own information exchanges, such as the Artnet Price Database \cite{artnet_nd}. Recognized artists with complete sets of known work, the Impressionists among them, represent a solved scarcity problem. As such, the price of limited-availability known work tends to remain stable in the market over time \cite{artTactic2013}, a fixed block of capital - “money on the walls” in the words of dealer Larry Gagosian \cite{keefe2023how}. When a given piece arrives at auction, the price it receives sets a value that may then shift the worth of an artist’s entire body of work.

When a painting is discovered that might be a fresh find from a registered and recognized artist, it is a moment of considerable interest and consequence. The catalog raisonné becomes crucial evidence in this authentication process. Because the real thing is worth so much, a work of art might come to seem more like a lottery ticket, and this kicks off the system in place to decide whether it is a forgery or an unknown treasure. The contents of the image might or might not matter to this process - but what does matter is the physical chemistry of the object, and then, the evidence that it has been documented in a chain of provenance, or historic ownership and registration through various institutions. Through this process, a painting is not so much created by an individual as made real by the recognition of appropriate accreditation. If it does not achieve the status of art, it lands in a different category: not art, an asset, but only an image, good for commerce and bad for capital gains. 

This system of value works closely with concepts developed in the mid-twentieth century, including Clement Greenberg’s “kitsch” \cite{greenberg1939avant} and Harold Bloom’s ‘anxiety of influence’ - the idea that artists must wrestle with their predecessors to create something genuinely new \cite{bloom1973anxiety}. Bloom argued that creative work always exists in dialogue with what came before, through what he termed 'poetic misprision' or creative misreading. Greenberg’s concept of ‘kitsch’ - cultural effects produced mechanically, without the underlying creative struggle of choice and skill - neatly fits the outputs of engines such as Stable Diffusion. The development of a unique style, then, a style that avoids kitsch, represents not just technical skill but the successful navigation of a very human anxiety - creating something recognizably new while acknowledging one's artistic lineage. Computational systems make this process yet more complex, as they absorb and reproduce stylistic elements without experiencing the psychological struggle both Greenberg and Bloom saw as central to artistic development.

The popularity of the light-and-color style of the Impressionists has given rise to many painters looking to mimic their output in a contemporary mode. If we examine these works, such as the output of Thomas Kinkade \cite{clapper2006thomas}\cite{bedell2011sentimental}, alongside work by generative AI systems, it seems clear that these are not so much art but themselves a form of style transfer. The scenes that are now produced in great volume as decoration look impressionistic, but they are not and cannot be Impressionist, a title reserved for the estates of a recognized body of former human people. Not quite a forgery and not quite an original, the work of Kinkade lives in history alongside paintings like those of Cornelius Krieghoff, a man earnestly dedicated to selling life in Quebec in 1850 by producing as many repetitive paintings of it as humanly possible \cite{krieghoff1850habitant}. Unlike the Impressionists, who were genuinely exploring the limits of a new technology, painters in this mode make a set product at a price-point accessible to people who would like to pay to look at it. Kinkade made a good living at it (\cite{hamrah2010cottage}, although not such a good death.

Style transfer for commercial decorative art is now being mechanized by generative systems.  It seems likely that such mechanization will increase over time, and that it will be perfected. The automation of style output will not have a meaningful impact on either the Impressionists - long-dead and converted to blue chip non-stocks, mainly resident in the vaults of various “high-value individuals” or state-level museums, outputting nothing new and protected as much as can be by international law on image rights \cite{musee_picasso_nd}. It will instead impact living artists, who have limited lifespans and - interestingly - styles that are unique, clear, and - potentially - novel. Such style is singular enough to be worth using to generate images that might be authored by an artist, but are not. This is the core complaint of the artists suing the groups behind Midjourney and Stable Diffusion \cite{goldstein2023b}. This means that, while creative styles across all disciplines might take a lifetime to develop in any practice - writing, code-making, music - the style of image-makers is a valuable target for computational approaches to production. This presents an intriguing role for HCI research:

How do we work to effectively recognize and report style once it has been trained? How can we build the accreditation for specific ways of seeing - a camera setting used mainly by a particular photographer of birds, a known way of handling digital ink specific to one comic artist - into our systems such that it cannot be avoided as generation and prompting become a type of creative practice in their own right?

The shift from careful documentation to mass ingestion represents more than a change in scale. Early computational approaches to art documentation, such as the Metropolitan Museum's first full digital push via the Innovaq system in 1983 \cite{lawrence1984museum}, and the Netherlands Institute's classification system, operated on what we might now call a 'human-centered' logic of curation. Each artwork was carefully considered, its attributes documented by experts who understood both its technical and cultural significance. This attention to provenance and attribution wasn't just scholarly pedantry - it was fundamental to establishing and maintaining the artwork's value in the market system. There are clear records discussing the challenge of style description for art catalogs, archive, and database systems as early as 1967, when the Netherlands Institute of Art History's pioneering work on computational art cataloging aimed to create what they called 'exact language' for describing artworks. When the Rijksbureau voor Kunsthistorische Documentatie developed their classification system in the 1950s, they included multiple entry points for each artwork, acknowledging that artistic influence operates across many dimensions. These early efforts at preservation and documentation have subsequently provided a bridge for the reproduction and transfer of the style itself.

Modern AI training sets represent an almost perfect inversion of this approach. Rather than careful curation, they operate through mass ingestion of images, often with minimal or incorrect attribution data \cite{crawford2021excavating}\cite{deng2009imagenet}\cite{torralba2011unbiased}. LAION-5B and similar datasets collapse the careful taxonomies developed by art historians into simple text-image pairs, treating a Renoir and a photograph of a Renoir reproduction as functionally equivalent. This flattening of attribution and provenance - the very elements that early computational systems were designed to preserve - creates a kind of stylistic hall of mirrors, where influence and originality become increasingly difficult to trace. The technical challenge for HCI, then, is not simply to reproduce style but to rebuild the crucial layers of attribution that earlier generations of archivists and technologists understood as essential to art's meaning and value.

If we are to follow Compton's logic of the “Bach faucet” as a system that produces an endless supply of a formerly-valuable cultural output, only to devalue it by so doing \cite{compton2023declaring}, there is a real risk that generative AI systems will spend a lot of energy producing something that is no better than spam. Computers are interesting to work with as art support because they do not - cannot - get bored. Yet this tirelessness is a double-edged sword: when art becomes less about individual creative expression in a unique moment or medium, and more about the repeated application of learned patterns, computers become problematic friends. Still, when properly constrained, systems of mechanical reproduction can be valuable tools - they don't suffer from repetitive strain injuries, their eyes don’t fog, and their components can be upgraded or replaced. The key challenge is ensuring that in becoming better tools, these systems don't subsume or replace the very human creativity they're meant to support. The machine must not eat the stylist it copies.

\section{A Proposal For Technical Intervention}
A modern system for documenting AI art's stylistic lineage might operate on multiple levels. At the training data level, it could maintain detailed attribution graphs showing which artists' works influenced specific model weights and how strongly. For instance, when an AI system generates an image in the style of Monet, the system could automatically generate a provenance document listing not just Monet's works in the training set, but also showing which specific paintings most heavily influenced the output's brush stroke patterns, color choices, and compositional elements. At the generation level, the system could track and document the series of prompts and parameters that led to the final image, preserving the human choices that shaped its creation. This documentation could be embedded within the image metadata itself via XMP or EXIF data, creating an unbroken chain of attribution from the original artists through to the final AI-assisted work. Such a system would serve not just legal or ethical purposes, but would help preserve the important connection between human creative development and computational reproduction.

\subsection{Current Approaches and Their Limitations}
Many technical systems papers related to AI within CHI and the broader community focus on communicating the provisions of AI tools to artists. This seems beside the point of art making, which is that once released, the tools are for artists to use, and use them they will. In this frame, the sometimes-embarrassing production of meme images is already successful at establishing the “art” of AI, more than any type of smart-infill background or color correction. These images are often not how computational researchers would like their work represented, and yet, along with misspellings of the word “strawberry,” they have vast power in public imagination \cite{jasanoff2015dreamscapes}.

This adversarial relationship offers us technical interventions from a different department of human-computer interaction: the computer and human security side. This offers a few practical cases for examination.

\subsection{Metadata Standards and Attribution}
The EXIF2Vec project attempts to use large language models to compare EXIF metadata and discover changes to the image via a pre-trained Word2Vec model. This work does not focus on image outputs at all, as it is primarily concerned with whether photographs have been tampered with or generated rather than produced by a camera at a place and time. It instead relies on semantic inconsistencies in the metadata to discover how images may have been altered in order to estimate correctness of description, and thus trustworthiness.

This framework offers a partial way forward for generated works: embed the data of where, when, and from what prompt the image was generated as part of the image standard for AI.

\begin{verbatim}
{
    "artist_id": "monet_claude",
    "artwork_id": "impression_sunrise_1872",
    "medium": "oil_on_canvas",
    "date_created": "1872",
    "style_elements": {
        "color_palette": [0.67, 0.45, 0.32],
        "brush_techniques": [
            "loose_brushwork",
            "visible_strokes"
        ],
        "compositional_elements": [
            "atmospheric_perspective",
            "natural_light"
        ],
        "subject_matter": [
            "landscape",
            "harbor",
            "sunrise"
        ]
    },
    "influence_weight": 0.85, 
    "consent_status": true,
    "usage_restrictions": [
        "attribution_required",
        "non_commercial_only",
        "not_in_france"
    ]
}
\end{verbatim}

\subsection{Visual Style Analysis}
XMP is the image data standard on most rich media formats. Designed to be used in an extensible way, the standard has moved from being text-only XML-like markup to storing complex data in a variety of formats. While XMP does not permit binary information encoding, it can store image data in Base64, and presumably could store other types of information in a cross-encoded format. This is recommended as a model for storing thumbnails, or in the case of Google’s image generation, depthmaps \cite{google2023depthmap}.

Following work on machine learning for games systems, such as AlphaGo, it seems possible to therefore encode a per-channel per-pixel set of predictive information that allows users to see which percentage of each pixel comes from each piece of the trained model outputting the image. This could then output a visual layer that comes with exploratory percentages, described visually as light-to-dark or some other axis.

Fiona Staples’ notable work with ink and color is here a good example, as her work has been used without her permission to train these systems \cite{staples2022instagram}, and is strikingly recognizable in other contexts. By applying x-y percentage layers to an image in XMP, we may be able to 'see' the edges of where one style begins and the next is applied. This also offers both creative and commercial possibilities: manipulate the graph form of the layer to push it more towards one style and away from another, and pay royalties for the most expensive styles.

\subsection{Verification and Validation}
Maintaining reversible and repeatable style transfer algorithms is the subject of considerable recent interest at CHI. Work on restorable arbitrary style transfer, style transfer guided by explicit aesthetic concerns, and scene classification for the fine arts have all seen recent publications, as has work in style transfer to written language. These efforts are in a good direction and offer promising routes forward in computational or quantitative understanding of the visual aspect of style transfer. They do not, as yet, offer ways to validate, document, and surface the processes that underlie the development of these styles, but by adding a layer that tracks and distills selection processes out of neural nets and into human-legible percentages, they offer a path to eventual validation processes. 

Recent work demonstrates several promising technical directions for style validation. Restorable arbitrary style transfer \cite{ma2024rast} techniques allow us to not only apply artistic styles but also reverse the process, providing a way to verify the relationship between original and generated works. This reversibility is important for documentation, as it allows us to trace the specific elements of style that have been transferred and potentially modified. Similarly, aesthetic-guided approaches like AesStyler \cite{yi2024aesstyler} explicitly model the relationship between style elements and aesthetic choices, making the transfer process more transparent and therefore more amenable to validation.

Scene classification \cite{huang2024scene} work in fine arts offers another avenue for validation by helping identify the specific contexts and elements that characterize an artist's style. This systematic approach to decomposing artistic elements provides a framework for verifying whether generated works maintain the essential characteristics of their source styles. Notably, similar approaches in natural language processing \cite{laquatra2024self} have already made significant progress in style attribution and verification, suggesting potential methods that could be adapted for visual art generation.

These validation approaches, when combined with our proposed documentation framework, offer a technical foundation for verifying style attribution that goes beyond simple visual similarity. By making the style transfer process both reversible and analyzable, we create opportunities for meaningful validation of artistic lineage.

\subsection{Evaluation Framework}
Developing such documentation systems requires us to fundamentally rethink evaluation frameworks. Where current assessments focus on output quality and style matching, a documentation-centered approach raises different questions. Systems must demonstrate their ability to trace stylistic elements to their human origins, document specific artistic influences, and preserve the context of when and how these elements emerged in human practice. Most importantly, they must record the sequence of human creative decisions that shaped their output.

Such metrics fundamentally shift system evaluation from questions of generation capability to questions of documentation fidelity. The key question becomes how meaningfully a system can document the human creative decisions and influences that shape its output. This approach acknowledges that the value of art lies not just in its aesthetic qualities but in its connection to finite human creative development—a connection that must be carefully preserved even as we develop more sophisticated means of reproduction.

\section{Conclusion}
The technical capacity to recognize computationally reproduced style demands a fundamental shift in how HCI approaches AI art. While the field has focused on improving generation capabilities and detection methods, we argue that HCI's true challenge lies in documenting and preserving the human creative decisions that give art its value. Where blockchain solutions have stalled on the promise of attribution, more meaningful technical interventions are not only possible but necessary. 

HCI researchers must develop attribution tracking systems that document not just which artists' works influenced an AI output, but how specific aesthetic elements were borrowed and transformed. We need metadata standards that embed complete stylistic provenance within generated works, and visualization tools that make these artistic influences legible to viewers. Most importantly, we must design interactive systems that give artists more control over their creative legacy in AI systems, documenting the human choices and iterations that shape each output. 

This is not merely a matter of proper citation or ethical practice. Rather, it represents an opportunity to document the development of human creative language across time, even as we move towards systems that can replicate these forms with increasing fidelity. The finite nature of human creative production - bounded by lifespan, by physical capacity, by the specificity of individual experience - is not a limitation to be overcome but the very foundation of artistic value. Each style represents something unrepeatable and worthy of documentation. Through this lens, HCI's role is not to perfect endless reproduction but to preserve and trace the human choices that make art meaningful in the first place. 

\begin{acks}
The authors would like to thank Dr. Alison McQueen at McMaster University for an authoritative view into what influence is and is not within art history. In addition, we would like to thank the faculty and staff of Sheridan College, Ontario, in particular Dani Sayeau and the illustration department for a view into the technical process of artistic style development.
\end{acks}

\bibliographystyle{ACM-Reference-Format}
\bibliography{references}

\end{document}